\documentclass{article}

\usepackage[preprint]{neurips_2025}

\usepackage[utf8]{inputenc} 
\usepackage[T1]{fontenc}    
\usepackage{hyperref}       
\usepackage{url}            
\usepackage{booktabs}       
\usepackage{amsfonts}       
\usepackage{nicefrac}       
\usepackage{microtype}      
\usepackage{xcolor}         

\usepackage{graphicx}
\usepackage{amssymb}
\usepackage{amsmath}
\usepackage{algorithm2e}
\SetKwComment{Comment}{/* }{ */}
\RestyleAlgo{ruled}

\newcommand{\red}[1]{\textcolor{red}{#1}}
\newcommand{\blue}[1]{\textcolor{blue}{#1}}

\usepackage{pifont}
\newcommand{\cmark}{\ding{51}}
\newcommand{\xmark}{\ding{55}}

\title{AGP: A Novel \textit{Arabidopsis thaliana} Genomics-Phenomics Dataset and its HyperGraph Baseline Benchmarking}

\author{
  \textbf{Manuel Serna-Aguilera}\textsuperscript{\texttt{1}}
  \textbf{Fiona L. Goggin}\textsuperscript{\texttt{2}} 
  \textbf{Aranyak Goswami}\textsuperscript{\texttt{3}} \\
  \textbf{Alexander Bucksch}\textsuperscript{\texttt{4}}
  \textbf{Suxing Liu}\textsuperscript{\texttt{4}}
  \textbf{Khoa Luu}\textsuperscript{\texttt{1}}\\
  \textsuperscript{1} Department of Electrical Engineering and Computer Science\\
  \textsuperscript{2} Department of Entomology and Plant Pathology, 
  \textsuperscript{3} Department of Animal Science\\
  \textbf{University of Arkansas, Fayetteville} \\
  \textsuperscript{4} School of Plant Sciences\\
  \textbf{University of Arizona, Tucson} \\
  \texttt{\{mserna, fgoggin, garanyak, khoaluu\}@uark.edu}\\
  \texttt{\{bucksch, suxingliu\}@arizona.edu}
}

\begin{document}

\maketitle

\begin{abstract}
    Understanding which genes control which traits in an organism remains one of the central challenges in biology. 
    Despite significant advances in data collection technology, our ability to map genes to traits is still limited.
    This genome-to-phenome (G2P) challenge spans several problem domains, including plant breeding, and requires models capable of reasoning over high-dimensional, heterogeneous, and biologically structured data.
    Currently, however, many datasets solely capture genetic information or solely capture phenotype information.
    Additionally, phenotype data is very heterogeneous, which many datasets do not fully capture. 
    The critical drawback is that these datasets are not integrated, that is, they do not link with each other to describe the same biological specimens. 
    This limits machine learning models' ability to be informed on the various aspects of these specimens, impacting the breadth of correlations learned, and therefore their ability to make more accurate predictions. 
    To address this gap, we present the \textit{\textbf{A}rabidopsis} \textbf{G}enomics-\textbf{P}henomics (\textbf{AGP}) Dataset, a curated multi-modal dataset linking gene expression profiles with phenotypic trait measurements in \textit{Arabidopsis thaliana}, a model organism in plant biology. 
    AGP supports tasks such as phenotype prediction and interpretable graph learning. 
    In addition, we benchmark conventional regression and explanatory baselines, including a biologically-informed hypergraph baseline, to validate gene-trait associations. To the best of our knowledge, this is the first dataset that provides multi-modal gene information and heterogeneous trait or phenotype data for the same \textit{Arabidopsis thaliana} specimens. With AGP, we aim to foster the research community towards accurately understanding the connection between genotypes and phenotypes using gene information, higher-order gene pairings, and trait data from several sources. 
\end{abstract}


\section{Introduction} \label{sec:intro}

Of the thousands of genes in an individual's genome and the hundreds of traits it displays, from their height to their health, which genes control which traits? The answers to this question are essential to nearly all applied life sciences, from crop improvement, animal breeding, to medical drug development. Unfortunately, our ability to provide answers is still quite limited due to data analytics limitations. Decoding the relationship between an organism’s genetic makeup, i.e., its genome, and its traits, i.e., its phenome--the total of its different phenotypes, requires identifying complex patterns that interlink multiple high-dimensional and heterogeneous datasets. Unfortunately, there is an absence of benchmarking data sets to facilitate the development of computational tools, particularly in the field of plant science. 
This hinders plant breeders’ efforts to meet increasing global food demands and combat emerging pests, diseases, and droughts.

\begin{table}[t]\centering
\caption{
    Comparison of publicly-available datasets or data repositories, which lack diverse, multi-omics components compared to our dataset--AGP. AGP includes gene-level and heterogeneous phenotype-level information, while other common datasets and repositories do not. A blue check \blue{\cmark} indicates the dataset contains the corresponding measurement type, while \red{\xmark} indicates otherwise.
}\label{tab:dataset-comparisons}
\scriptsize 
\begin{tabular}{lcccccc}
\textbf{} &\textbf{} &\textbf{Image-derived} &\textbf{Manual Measures of} &\textbf{Photosynthetic} &\textbf{Gene} \\
\textbf{Dataset/Repository} &\textbf{Images} &\textbf{Phenotypes} &\textbf{Growth/Development} &\textbf{Measurements} &\textbf{Expression} \\\midrule
I-PPD \cite{minervini-2016, minervini-dataset-2015} &\blue{\cmark} &\blue{\cmark} &\red{\xmark} &\red{\xmark} &\red{\xmark} \\
SAD \cite{ward-moghadam-dataset-2018, ward2019deepleafsegmentationusing} &\blue{\cmark} &\blue{\cmark} &\red{\xmark} &\red{\xmark} &\red{\xmark} \\
UNL-PPD \cite{unl-ppd} &\blue{\cmark} &\blue{\cmark} &\red{\xmark} &\red{\xmark} &\red{\xmark} \\
araPheno \cite{arapheno-dataset-site, arapheno-seren-2017} &\red{\xmark} &\red{\xmark} &\blue{\cmark} &\red{\xmark} &\red{\xmark} \\
Photosynq \cite{photosynq-dataset-site} &\red{\xmark} &\red{\xmark} &\red{\xmark} &\blue{\cmark} &\red{\xmark} \\
NCBI-GEO-SRA \cite{geo-repository-site, sra-site, ncbi-repository-2002} &\red{\xmark} &\red{\xmark} &\red{\xmark} &\red{\xmark} &\blue{\cmark} \\
ENA \cite{sra-site} &\red{\xmark} &\red{\xmark} &\red{\xmark} &\red{\xmark} &\blue{\cmark} \\
TAIR \cite{tair-site, tair-paper-2001, tallon-rnaseq-benchmarking-2024} &\red{\xmark} &\red{\xmark} &\red{\xmark} &\red{\xmark} &\blue{\cmark} \\
\midrule
\textbf{AGP} (Ours) &\blue{\cmark} &\blue{\cmark} &\blue{\cmark} &\blue{\cmark} &\blue{\cmark} \\
\bottomrule
\end{tabular}
\end{table}

\noindent
\textbf{Limitations in Prior Work.} Research to crack the ``genome-to-phenome'' (G2P) challenge increasingly relies on high-dimensional and heterogeneous data to capture as many different aspects as possible of an individual specimen’s genome and phenome. 
Genomics in part provides us with gene expression measurements or transcriptomics profiles, which provide information about when and where each gene in that genome is expressed. 
Plant phenomic data typically has fewer features but is highly heterogenous, ranging from images of shape and size to manual observations of development to spectrophotometric measurements of processes like photosynthesis. Combining data from more than one ``omics'' approach (i.e., multi-omics) is more effective at linking genes to specific phenotypes than any single omics approach alone \cite{minervini-2016, minervini-dataset-2015, ward-moghadam-dataset-2018, ward2019deepleafsegmentationusing, unl-ppd, arapheno-dataset-site, arapheno-seren-2017, photosynq-dataset-site}. Despite this, different types of plant omics data are siloed in separate repositories as shown in Table \ref{tab:dataset-comparisons}, and there is a lack of comprehensive datasets that combine genomic and phenomic profiles from the same individuals to allow correlative analyses. Due to this lack of benchmarking data, the machine learning community has not kept up with the challenges of analyzing multi-omics data. Although there is much discussion in the biology literature concerning an artificial intelligence model that could understand correlations between samples in omics data \cite{cavill-omics-2015, cembrowska-omics-2023, demidchik-phenomics-2020, yang-omics-review-2021, gao-omics-short-review-2023, depuydt-omics-paper-2023, yan-ml-omics-review-2023, zhang-multi-omics-2022, mohammed-ai-omics-2023, wenhui-ml-omics-2024}, no such actual models seem to exist. Therefore, there is a critical need for a biologically-informed model that can map gene expression to heterogeneous phenotypes while allowing investigations into its explainability.

\noindent
\textbf{Problem Motivation.} To address gaps and limitations in the machine learning space, we present the \textit{Arabidopsis} Genomics-Phenomics (AGP) dataset with genomics and phenomics data linked to several specimens of the model plant \textit{Arabidopsis thaliana}. Compared to other datasets, as in Table \ref{tab:dataset-comparisons}, we combine gene expression and heterogeneous phenotype measurements with respect to the same plant specimens. The multi-modal gene-level data contains more than transcriptomic profiles; we also provide text summaries, sequences, and unique identifiers for each gene. To build biologically-informed baselines, we utilize hypergraphs to connect genes via higher-order relationships. This allows hypergraphs to capture and encode complex structures found in nature, similar to past works into data with higher-order relationships \cite{yadati2019hypergcnnewmethodtraining, yadati2019hypergcnnewmethodtraining, wang2021heterogeneousgraphattentionnetwork, zhang-aminer-2019, wu20153dshapenetsdeeprepresentation, wang2021heterogeneousgraphattentionnetwork, twenty_newsgroups_113, chien2022allsetmultisetfunctionframework}. 
Combined with deep learning, such data structures have succeeded in different scenarios in recent years. Hypergraph methods have been subject to study spanning decades \cite{hwang-hypergraph-genes-2008, gao-social-media-2013, zien-1999, zhou-neurips2006, rodriguez2003, Bolla-1993, alistarh-neurips2015, li2017inhomogeneoushypergraphclusteringapplications} and have seen much success when combined with deep learning \cite{feng2019hypergraphneuralnetworks, bai2020hypergraphconvolutionhypergraphattention, yadati2019hypergcnnewmethodtraining, chen2023hytrelhypergraphenhancedtabulardata, yadati-nerips2020, zheng2024cursorscalablemixedorderhypergraph, fan-cvpr2024, kim-cvpr2020, khan2023learningsituationhypergraphsvideo, wei2022augmentationshypergraphcontrastivelearning, wang2023equivarianthypergraphdiffusionneural, wang2023hypergraphenergyfunctionshypergraph, kim2024hypeboygenerativeselfsupervisedrepresentation}. We then provide baseline benchmark results for regression and, importantly, explanatory baselines using SHAP \cite{lundberg-2017-shap-paper}, and offer insight into the limitations of current works.

\noindent
\textbf{Contributions of this Work.} We summarize the contributions of this work below. 
\begin{itemize}
    \item We contribute the novel AGP dataset, which contains multi-modal data for over 27,000 expressed genes. AGP also provides heterogeneous phenotype measurements for the same specimens. 
    \item We establish baseline regression and naive explainability results on both naive and baseline biologically-informed models.
    \item We provide insight into the explainability of the baseline methods, and discuss limitations to foster future research into biologically explaining how genotypes influence organism traits.
\end{itemize}

\section{Background and Related Work} \label{sec:background-related-work}

\begin{figure*}
    \centering
    \includegraphics[width=1.0\linewidth]{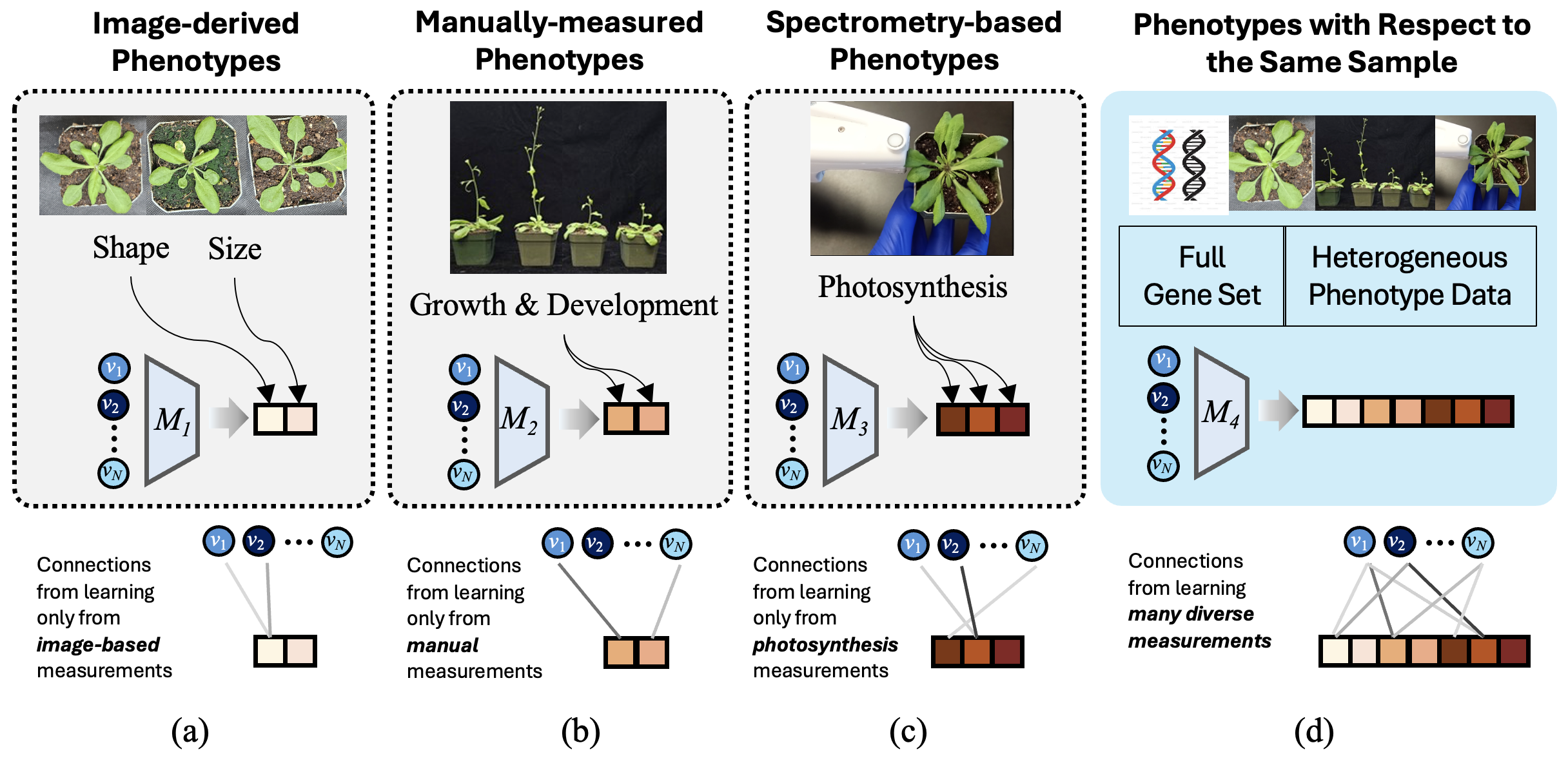}
    \caption{
        An illustration of the problem non-heterogeneous phenomics data poses to downstream tasks for the \textit{Arabidopsis} model plant. 
        \textbf{(a)} Model $M_1$ is trained only on image-derived phenotype data to predict phenotypes like leaf area, given genes $v_1$ up to $v_N$. Such models suffer from limited perspective, capturing a small portion of real relationships between genes and traits. 
        \textbf{(b)} Model $M_2$ is trained only on manual measurements to predict growth and development, but has no way to access $M_1$'s knowledge, and suffers from the same problem.
        \textbf{(c)} Model $M_3$ is trained only on spectrometry data and suffers the same lack of perspective. 
        \textbf{(d)} With the breadth of phenotypes provided by AGP, models like $M_4$ learn to correlate genes to multiple traits of the specimens.
    }
    \label{fig:our-phenotype-data}
\end{figure*}





\subsection{The Model Plant \textit{Arabidopsis thaliana}}
To develop a benchmark dataset for multi-omics data analysis, we chose the thale cress plant (\textit{Arabidopsis thaliana}). 
``Omics'' or ``multi-omics'' data refers to measurements from different biological systems in an organism \cite{cavill-omics-2015, cembrowska-omics-2023}.
Much like rats and mice are ``model species'' that can help us learn about human health, the thale cress is a model plant that is widely used to learn about important crops like corn and rice \cite{Woodward2018-hp}. The relatively small genome and short lifecycle of the thale cress have enabled much progress in studying its biological processes. 
Of the many different types of thale cress that are available to study, we chose to focus on four lines or genotypes, with each line having specific alterations to some genes, i.e., mutations that give rise to different traits. These lines show important differences in their shape, photosynthetic efficiency, and speed of development--characteristics that are all important for plant breeding. Thus, these lines provide a good test bed to combine omics approaches with hypergraphs to explore the influence of the transcriptome on important plant phenotypes. Further details are provided in the supplementary materials.

\subsection{Multi-Omics Technologies}
Recent advancements in biology have led to a surge in multi-omics data analysis \cite{cavill-omics-2015, cembrowska-omics-2023}, where different biological systems—such as the genome (genomics) and transcriptome (transcriptomics)—are studied in parallel to understand their influence on an organism’s phenome, or observable traits. 
The genome of the particular specimen (in our case, an individual thale cress plant) consists of the complete set of genes in that individual.
Genomics is the study of the structure, function, and evolution of genomes within a given organism or community of organisms \cite{omics-definitions-paper-heavey-2022}. 
Transcriptomics is concerned with gene expression patterns, and nowadays RNA sequencing, or RNAseq analysis, allows us to quantify the expression of all or nearly all the genes in the genome at one time. \cite{mansoor-2025}.
This gives researchers insight into what genes are triggered during developmental stages or certain environmental stimuli. 
All these biological systems affect an organism’s phenome, which can be observed by image analysis \cite{minervini-2016} (e.g., leaf area, color analysis), manual measurements (e.g., recording growth rates), and via more advanced tools such as spectrometry (e.g., to understand photosynthesis rates). 
In our work, we specifically focus on integrating or linking gene expression data (via RNA-seq) with heterogeneous phenotypic measurements, including manual traits (e.g., flower development), chemical data (e.g., photosynthetic capacity), and image-derived traits, as shown in Figure \ref{fig:our-phenotype-data} as opposed to other methods that process this data in isolation.
Cembrowska et al. \cite{cembrowska-omics-2023} and past surveys \cite{demidchik-phenomics-2020, yang-omics-review-2021, gao-omics-short-review-2023, depuydt-omics-paper-2023, yan-ml-omics-review-2023, zhang-multi-omics-2022, mohammed-ai-omics-2023, wenhui-ml-omics-2024} provide more insight into omics research.

\subsection{Multi-Omics Dataset Landscape}
While high-throughput omics data collection has improved considerably, most existing datasets fail to provide linked multi-omics measurements for the same specimens. 
Many recent works, therefore, operate on this limited data \cite{inferring-phenotypes-cheng-2021, flood-2016-circadian-rhythms, ubbens-deep-plant-phenomics-2017}. 
As shown in Table \ref{tab:dataset-comparisons}, current repositories often contain either image-based phenotypes or gene expression data, but rarely both. 
Moreover, many studies apply gene-thresholding techniques that exclude large portions of the genome due to computational constraints. 
As a result, even if recent works tackle G2P, they unfortunately run into computational or runtime limitations if given a complete set of genes. 
It is the case not just for thale cress-related research \cite{tian-2007-phenomics, barreda-et-al}, but also with human data in the medical field \cite{rahimikollu-slide-2024}. 
Our dataset addresses this gap by offering complete gene expression profiles paired with rich phenotypic traits, enabling direct genotype-to-phenotype modeling. 
This positions our dataset as a valuable benchmark for developing scalable, integrative machine learning methods.
Detailed descriptions of these datasets from Table \ref{tab:dataset-comparisons} are provided in the supplementary materials.

\subsection{Hypergraphs and Higher-Order Relations} 
Hypergraphs offer a compact manner to take large genomes beyond pair-wise groupings \cite{berge-hyper-textbook-1989}. Hwang et al. \cite{hwang-hypergraph-genes-2008} investigated hypergraphs for connecting gene expressions and protein interactions; this is distinct from our work, which focuses on gene-level and phenotype connections. Hypergraph research has seen much development over the decades \cite{zien-1999, rodriguez2003, zhou-neurips2006, yadati-nerips2020}. 
Zhou et al. \cite{zhou-neurips2006} introduced a hypergraph clique expansion and grouped edges to form hyperedges for the classification task. In the past decade, deep learning has become integrated into hypergraph tasks, formulating learnable hypergraph convolution \cite{yadati2019hypergcnnewmethodtraining, feng2019hypergraphneuralnetworks}, hypergraph attention with Bai et al. \cite{bai2020hypergraphconvolutionhypergraphattention}, and deeper graph networks with the work of Huang et al \cite{huang2021unignnunifiedframeworkgraph}.
Hypergraphs with deep learning have been used in many tasks. Among hot topics include scene generation built on graphs and hypergraphs \cite{nguyen2025hyperglmhypergraphvideoscene, }, vision-natural language scenarios \cite{khan2023learningsituationhypergraphsvideo, kim-cvpr2020}, federated learning \cite{fan-cvpr2024}, recursive hyper edges \cite{yadati-nerips2020}, hypergraph matching \cite{zheng2024cursorscalablemixedorderhypergraph}, contrastive learning \cite{wei2022augmentationshypergraphcontrastivelearning}, and tabular data \cite{chen2023hytrelhypergraphenhancedtabulardata}. The explicit consideration of hyperedge weights is the subject of other works \cite{gao-social-media-2013, hwang-hypergraph-genes-2008}, and further groupings such as clustering or partitioning of hypergraphs, to maximize similarity within the groupings, have also been researched over the years \cite{zien-1999, Bolla-1993, alistarh-neurips2015, li2017inhomogeneoushypergraphclusteringapplications}. Common tasks such as vertex or node classification and hyperedge prediction have been the subject of much recent work such as the works of Kim et al. \cite{kim2024hypeboygenerativeselfsupervisedrepresentation}, Wang \cite{wang2023hypergraphenergyfunctionshypergraph}, and Wang \cite{wang2023equivarianthypergraphdiffusionneural}. Hot topics, including large language models, have seen increasing integration of hypergraphs to address hallucination \cite{feng2024graphslargelanguagemodels, chu2024llmguidedmultiviewhypergraphlearning, huang2025hyperghypergraphenhancedllmsstructured}. The long tail problem and hypergraphs has also seen recent research, from long tail visual input \cite{HAN2024112400}, recommender systems (for hypergraphs) \cite{han2022searchbehaviorpredictionhypergraph}, and finally, combining knowledge graphs, hypergraphs, and long-tail representations \cite{liu2023selfsuperviseddynamichypergraphrecommendation}. Thus, this work provides a strong foundation for applying hypergraph models to the G2P challenge.


\section{The Proposed AGP Dataset} \label{sec:dataset}
In this section, we provide an overview of AGP. It contains gene data of multiple modalities, image-derived phenotype data, manually-collected data, and spectrometry-based data. We first discuss the data collection process in Section \ref{subsec:data-collection}, followed by overviews of the gene-level data in Section \ref{subsec:gene-level-info}, how we connect the genes to form hyperedges in Section \ref{subsec:build-hyperedges}, and finally, an overview of the phenotype data in Section \ref{subsec:phenotype-level-info}.

\subsection{Data Collection} \label{subsec:data-collection}
This study utilized four types of thale cress (so-called lines or genotypes) that differ in shape, developmental rates, and photosynthetic efficiency. These differences are caused by mutations in the genes. In order to quantify these phenotypic differences and also to provide examples of the heterogenous data types that are common in plant phenomics, we use three different phenotype measurements to characterize the plants: image-based measurements of shape and size, manual measurements of reproductive development, and spectrophotometric measurements of different aspects of photosynthesis (Figure \ref{fig:our-phenotype-data}).  Our spectrophotometric measurements also provided us with information about light and temperature, important environmental conditions that can influence the relationship between genomes and phenomes. After taking these measurements, we destructively sampled a subset of the plants to study their gene expression profiles using RNAseq analysis. The remaining plants that were not harvested for RNAseq analysis were allowed to complete their life cycle in order to manually measure seed production, another important phenotype for both plant breeding and evolutionary studies. 

\subsection{Gene-Level Information} \label{subsec:gene-level-info}


\begin{figure*}
    \centering
    \includegraphics[width=1.0\linewidth]{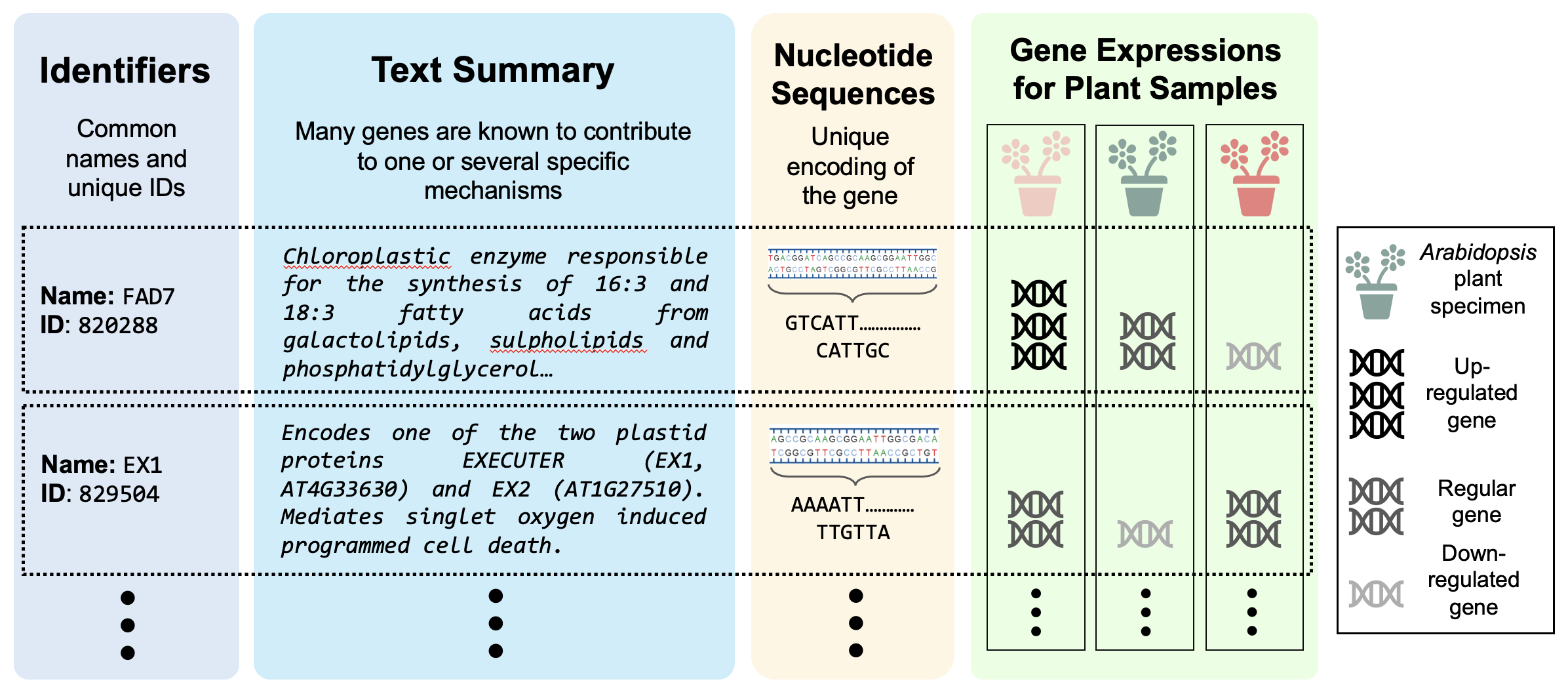}
    \caption{
        A visualization of the different features for the gene-level data provided by AGP. Genes have identifiers, text descriptions indicating what functions they influence. We also have the unique nucleotide sequence for each gene--its blueprint encoded by a string of characters. Finally, we add our gene expression data for all genes across different plant specimens. 
    }
    \label{fig:gene-information}
\end{figure*}

\textbf{Transcriptomics}. The AGP dataset contains gene expression or transcriptomics data for over 27,000 protein coding genes. 
It results from subjecting 24 plant specimens to RNAseq analysis, giving us direct measurements of how much each gene in the genome is expressed. 
As shown in Figure \ref{fig:gene-information}, different specimens show different patterns in gene expression due to being from different lines.

\textbf{Multimodal Gene Information}. In addition to RNAseq data, the AGP dataset contains additional information for each gene of Arabidopsis, extracted and compiled from the National Center for Biological Information (NCBI\footnote{https://www.ncbi.nlm.nih.gov/datasets/gene/taxon/3702/}). Each gene has identifiers associated with it, these are IDs given to by NCBI, as well as common names (e.g., EX1). The most commonly-studied genes have an associated text summary, giving some details as to what processes it is thought to influence. We organize all of this gene-level data such it can be queried in a local dataframe, giving users a quick reference for all genes of interest and quickly ascertain what the queried genes do. 

\subsection{Connecting Genes for Biologically-Informed Models} \label{subsec:build-hyperedges}

\begin{figure*}
    \centering
    \includegraphics[width=1.0\linewidth]{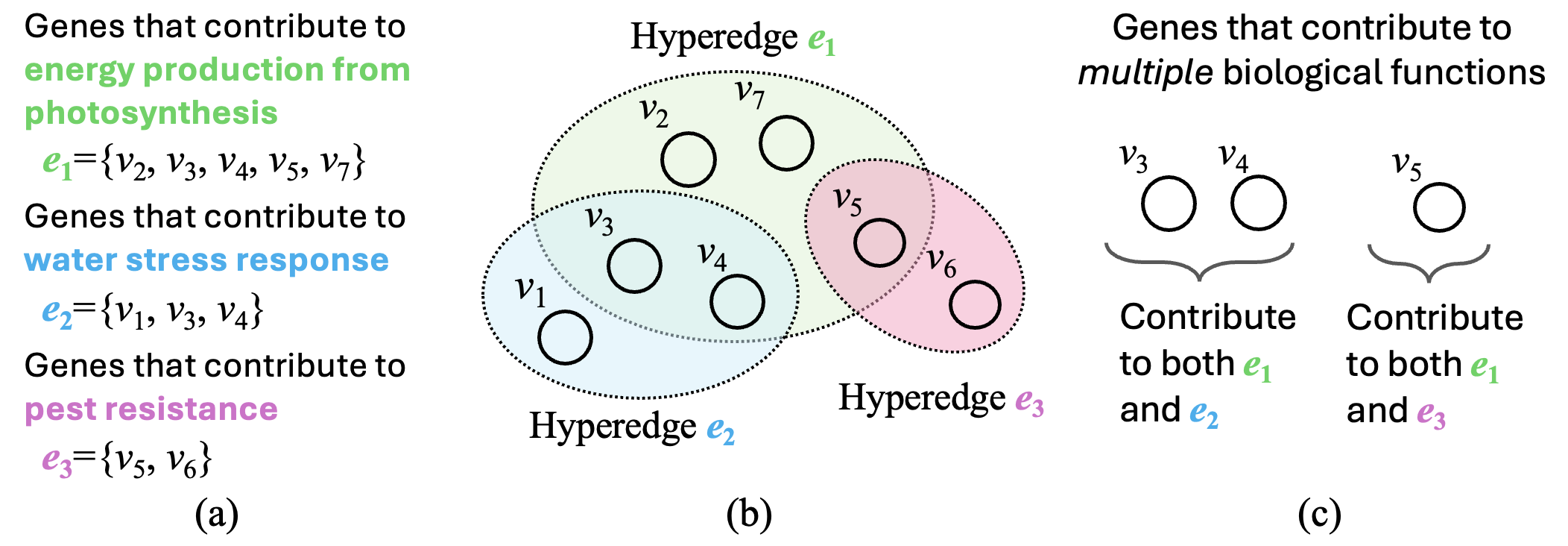}
    \caption{
        A visualization of translating the biological functions of the genes the thale cress to higher-order pairings.
        \textbf{(a)} Thousands of functions exist for many systems in lifeforms, each containing genes that regulate those functions. 
        \textbf{(b)} The sets of genes can form hyperedges.
        \textbf{(c)} Hyperedges are connected by overlapping genes.
        Best viewed in color.
    }
    \label{fig:hyperedge-relations}
\end{figure*}

To connect tens of thousands of genes, we collect biological function information from the Gene Ontology (GO) repository\footnote{https://geneontology.org/}. These functions have unique identifiers similar to genes. Each function has an identifier, a text summary, and most importantly, relations to which genes contribute. 
For instance, the biological function identified as \texttt{GO:0047484} has the description: \text{``regulation of response to osmotic stress’’}, and we can pull up the genes known to contribute to that function. The scope of each GO term can vary from a broad function to a specific function, making the structure between GO terms somewhat nested; we have over 7,000 of these functions.
We observe that sets of genes can represent these biological functions. Therefore, we can use hyperedges to encode these sets as shown in Figure \ref{fig:hyperedge-relations}.

\subsection{Phenotype-Level Information} \label{subsec:phenotype-level-info}
We have collected phenotype measurements or “parameters” for 77 physical plant specimens from four different lines (number of specimens for each described in Supplementary). A total of 41 parameters were measured, but for the purposes of benchmarking, we focused on five parameters of interest. These five parameters are: \textbf{(1)} Solidity: a measure of how compact or “bushy” the plant is (based on image analysis); \textbf{(2)} Height of the inflorescence or flower stalk: an indicator of how far into reproductive development a plant is (measured manually); \textbf{(3)} FvP/FmP: a measure of how efficiently the plant can channel light energy into photosynthesis (collected with a spectrometer); 
\textbf{(4)} qL: a measure of the chemical state of an important compound called plastoquinone in photosynthesis (also collected with a spectrometer); 
and \textbf{(5)} Leaf temperature differential: the difference in temperature between a leaf and its surroundings (also collected with a spectrometer). 
These parameters were chosen because they represent each of the heterogenous data types in our data set, and because they show important variation among our four different thale cress lines. A subset of the plants used for phenotype measurement (6 plants per thale cress line) were also used for RNAseq analysis, so we have paired phenotypic and genomic data for a total of 24 plants. 

\section{Methodology} \label{sec:methodology}

In this section, we give an overview our baseline method, i.e., \textbf{H}igher-order \textbf{B}iologically-informed \textbf{F}ramework (\textbf{HBF}). We first cover preliminaries in Section \ref{subsec:background}, then provide details on data preprocessing in Section \ref{subsec:data-preprocessing}, and give an overview of HBF in Section \ref{subsec:framework}.

\subsection{Hypergraph Definitions and Notations} \label{subsec:background}
We use the hypergraph as the backbone model for our HBF evaluations. A hypergraph is defined as  $\mathcal{G} = \{\mathcal{V}, \mathcal{E}, \textbf{W}\}$. Let $\mathcal{V}$ be the set of vertices (or nodes) and a particular vertex $v \in \mathcal{V}$. We denote $\mathcal{E}$ as a set of hyperedges, and each element $e \in \mathcal{E}$ is a set of vertices. Finally, let $\textbf{W} \in \mathbb{R}^{|\mathcal{E}| \times |\mathcal{E}|}$ be a diagonal matrix containing the hyperedge weights, where each diagonal element $\textbf{W}_{e,e} \in \mathbb{R}$ ($e$ can index a row or column). Furthermore, the hypergraph incidence matrix $\textbf{H} \in \mathbb{R}^{|\mathcal{V}| \times |\mathcal{E}|}$ is a binary matrix that gives whether or not a vertex $v$ is in a particular hyperedge $e$, with $\textbf{H}_{v,e} = 0$ if $v \in e$, else $\textbf{H}_{v,e} = 1$ if $v \notin e$.
Let $\textbf{B} \in \mathbb{R}^{|\mathcal{E}| \times |\mathcal{E}|}$ be a diagonal matrix for the degree of each hyperedge, and let $\textbf{D} \in \mathbb{R}^{|\mathcal{V}| \times |\mathcal{V}|}$ be a diagonal matrix for the degree of each vertex.
Thus, the degree of a hyperedge is defined as $\textbf{D}_{e,e} = \Sigma_{e \in \mathcal{E}} \textbf{W}_{e,e}\textbf{H}_{v,e}$ and the degree of a vertex $\textbf{B}_{v,v} = \Sigma_{v \in \mathcal{V}} \textbf{H}_{v,e}$. Following \cite{bai2020hypergraphconvolutionhypergraphattention, feng2019hypergraphneuralnetworks}, let the standard hypergraph convolution be defined as in Eqn. \eqref{eq:hypergraph-conv-bai}.
\begin{equation} \label{eq:hypergraph-conv-bai}
    \textbf{X}^{(l+1)} = \sigma\Big(\textbf{D}^{-1}\textbf{H}\textbf{W}\textbf{B}^{-1}\textbf{H}^{\mathsf{T}}\textbf{X}^{(l)}\textbf{P}\Big)
\end{equation}
\noindent
The learnable weights are denoted by $\textbf{P} \in \mathbb{R}^{F_{1} \times F_{2}}$. The matrix $\textbf{H}^{\mathsf{T}}$ is the transpose of \textbf{H}, and other notations are the same as in the simple graph formulation.
The input, the gene expression features for a particular specimen, is denoted as $\textbf{X}$, a column vector of length $N_G$, where $G$ is the number of genes with expression values.

\subsection{Data Preprocessing} \label{subsec:data-preprocessing}
\textbf{Gene-Level Data.} The gene expression data comes in two forms. First is the raw gene expression values, but we do not use these as this data does not account for differences between the plant specimens, which affect how much a gene will be expressed. The second form normalizes the raw values and is referred to as Fragments Per Kilobase of transcript per Million mapped reads (FPKM). From the FPKM values, we may also compute Transcripts per Million (TPM), another popular normalization. These values make up the gene node features $\textbf{X}$, where the dimensionality for each node may range from 1 (if we want to process one plant at a time) up to multiple plants' measurements. $\textbf{X}$ is then standardized with respect to the plant specimens to preserve variance within the same sample.

\textbf{Phenotype-Level Data.} 
In this work, we focus on $\rho = 5$ phenotype parameters across $N_P=23$ \textit{Arabidopsis} plant specimens. Let $P$ refer to the number of plant specimens while $\rho$ refers to the number of phenotypes. Generally, the phenotype targets are denoted as $\textbf{Y} \in \mathbb{R}^{N_P \times \rho}$ or $\textbf{Y} \in \mathbb{R}^{N_P \times 1}$ when focusing on one phenotype. We also standardize the measurements in \textbf{Y} with respect to the samples to maintain the variance among measurements of the sample specimen.

\subsection{Baseline Framework} \label{subsec:framework}

\begin{figure*}
    \centering
    \includegraphics[width=0.8\linewidth]{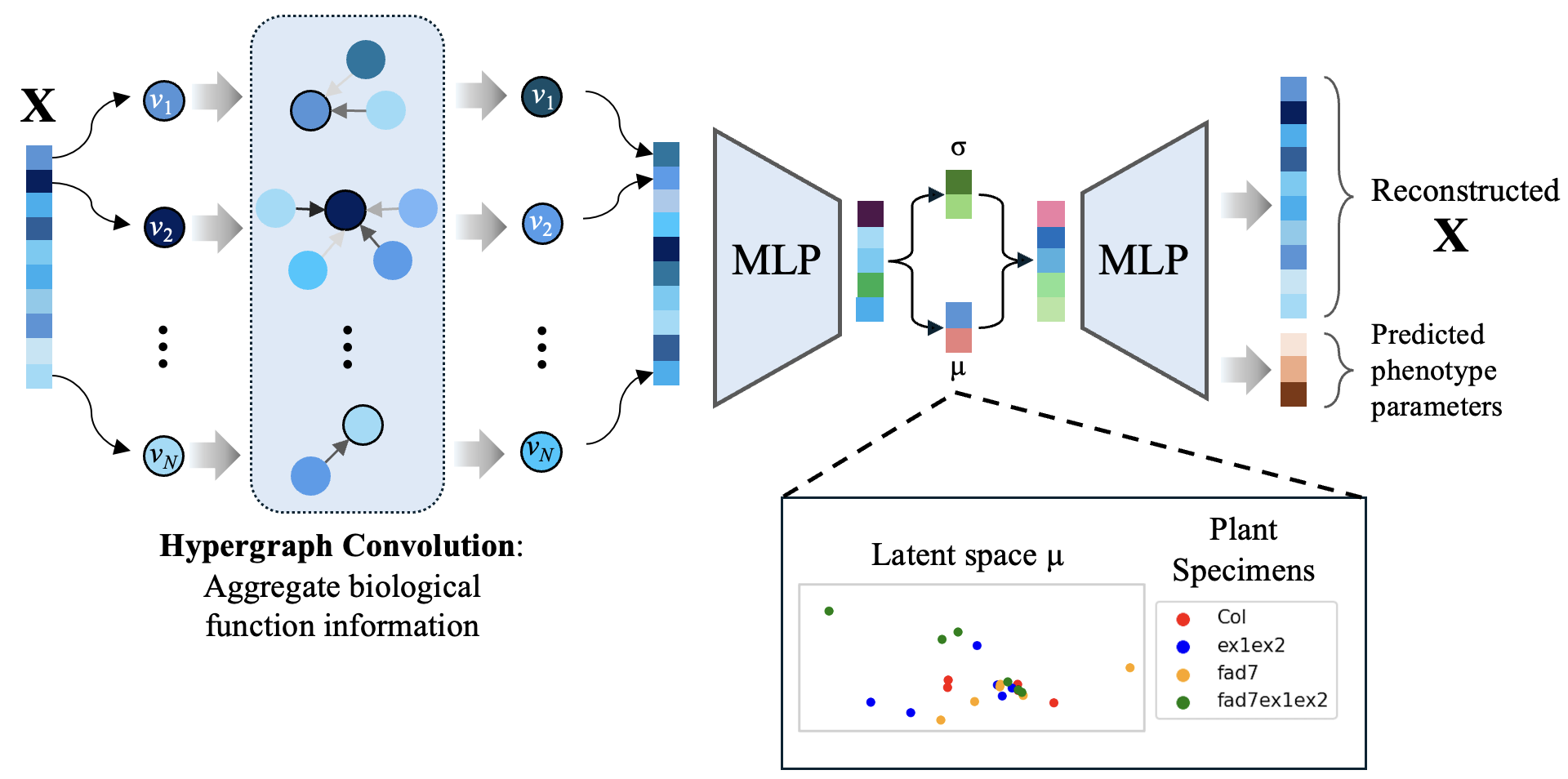}
    \caption{
        The baseline HBF, a VAE for gene feature reconstruction and regression. The gene features or the transcriptomics data, for over 27,000 genes, are passed on to the higher-dimensional hypergraph layer. We then compress these aggregated gene interactions into a latent space, giving us insight into how thale cress plant specimens could be related.
    }
    \label{fig:genomics-phenomics-frameworks}
\end{figure*}

    
    
        
        
        
        
        
    

The input is denoted as $\textbf{X} \in \mathbb{R}^{N_P \times N_g}$ ($g$ refers to genes), and the outputs $\hat{\textbf{Y}} \in \mathbb{R}^{N_p \times \rho}$ are the predictions of the phenotype parameters, and the ground truths \textbf{Y} are the same shape. We first train a random forest model for regression. The second model we train is a straightforward Multilayer Perceptron (MLP) with steadily decreasing dimensions. The third model is the same base MLP model, but with added hypergraph convolution layers (Equation \ref{eq:hypergraph-conv-bai}) where the single gene expression is expanded into a higher-dimension space (e.g., 64). The forward- and back-propagation are done with respect to one plant sample, that is, the batch size is 1, so the hypergraph layers can learn one plant's connections at a time. 
While the models are optimized for regression, we are more interested in investigating how current \textit{explainability} tools could help us understand how genes (their expression) and phenotypes \textit{correlate}. 
In other words, we wish to know which genes have the greatest \textit{correlations} with respect to a phenotype.
Therefore, we use SHAP \cite{lundberg-2017-shap-paper} as the explanatory baseline for all of the models we evaluate, including the HBF models; we compute the SHAP values for each input feature with respect to each phenotype.

\section{Experimental Results and Discussion} \label{sec:experiments}

In this section, we first review the experimental setup, including what experiments are conducted, in Section \ref{subsec:experiment-setup}. Next in Section \ref{subsec:regression-experiments} we discuss the regression experiments with respect to Table \ref{tab:regression-metrics}. Finally, in Section \ref{subec:explainability-baselines}, we discuss the explanatory genes found in Tables \ref{tab:top-genes-12} and \ref{tab:top-genes-345}.

\subsection{Experiment Setup} \label{subsec:experiment-setup}
In this section, we provide further implementation details in the experiments. The baseline experiments can be performed on a machine with an RTX 3060 with approximately 12 GB of memory. 
The random forest regressor was implemented with scikit-learn \cite{scikit-learn}, and all other models with Pytorch \cite{pytorch} and Pytorch Geometric layers\footnote{\url{https://pytorch-geometric.readthedocs.io/en/latest/index.html}}. With respect to explainability, we use different explainer objects from SHAP due to the different nature of each model. For random forest, we use the \texttt{TreeExplainer} class, while for MLP, we can use the \text{DeepExplainer} class. Due to hypergraphs necessitating their explanatory frameworks, unfortunately, no current open-source implementation exists \cite{su2024explaininghypergraphneuralnetworks}. To include hypergraphs in this specific problem setting, we used the \texttt{KernelExplainer} due to the constraint of processing one plant specimen's data at a time.

\subsection{Hypergraph Regression} \label{subsec:regression-experiments}


\begin{table}[!t]\centering
\caption{
    The Mean Absolute Error (MAE) and root Mean Square Error (rMSE) for Regression for naive and biologically-informed baselines, including the baseline VAE. We average metrics on all random test set splits and report the standard deviation (stdev). Model names with hypergraph layers are shown in \textbf{bold}.
}\label{tab:regression-metrics}
\scriptsize
\begin{tabular}{lccccc}\toprule
\textbf{Model} &\textbf{Average MAE} &\textbf{Median MAE} &\textbf{Average rMSE} &\textbf{Median rMSE} \\
\midrule
Random Forest &0.5768 $\pm$ 0.0519 &0.5696 &0.6578 $\pm$ 0.0882 &0.6617 \\\
MLP &0.8441 $\pm$ 0.3915 &0.5842 &1.3691 $\pm$ 0.9946 &0.6986 \\
\textbf{HGNN + MLP} &1.2612 $\pm$ 0.9791 &0.5883 &2.5352 $\pm$ 2.6467 &0.6957 \\
\textbf{HGNN + VAE} &0.5530 $\pm$ 0.0568 &0.531 &0.6354 $\pm$ 0.102 &0.5907 \\
\bottomrule
\end{tabular}
\end{table}


\begin{table}[!t]\centering
\caption{Top explanatory genes with respect to Solidity and Influorescence Height. Model names with hypergraph layers are shown in \textbf{bold}.}\label{tab:top-genes-12}
\scriptsize
\begin{tabular}{lrrrrr}
\textbf{} &\multicolumn{2}{c}{\textbf{Solidity}} &\multicolumn{2}{c}{\textbf{Influorescence Height}} \\\cmidrule{2-5}
\textbf{Model} &\textbf{SHAP Value} &\textbf{Gene Name} &\textbf{SHAP Value} &\textbf{Gene Name} \\\midrule
Random &0.1193 &AT2G01100 &0.0512 &AT2G01100 \\
Forest &0.0762 &PDE331 &0.0356 &PDE331 \\
&0.0325 &EXGT-A3 &0.0325 &PAHX \\\midrule
MLP &0.011 &CRK26 &0.0131 &QPT \\
&0.0107 &AT5G46770 &0.0104 &LHCA2 \\
&0.0104 &AT5G35069 &0.0101 &PSAL \\\midrule
\textbf{HGNN} &> 1.0 &AT3G13672 &0.4074 &AT4G00413 \\
\textbf{+ MLP} &> 1.0 &AT3G29720 &0.2745 &DHAR1 \\
&> 1.0 &AT5G37280 &0.1898 &MCM8 \\
\bottomrule
\end{tabular}
\end{table}

\begin{table}[!t]\centering
\caption{Top explanatory genes with respect to the spectrometry-based phenotype measurements. Model names with hypergraph layers are shown in \textbf{bold}.}\label{tab:top-genes-345}
\scriptsize
\begin{tabular}{lrrrrrrr}
&\multicolumn{2}{c}{\textbf{qL}} &\multicolumn{2}{c}{\textbf{FvP/FmP}} &\multicolumn{2}{c}{\textbf{Leaf Temp. Differential}} \\\cmidrule{2-7}
\textbf{Model} &\textbf{SHAP Value} &\textbf{Gene Name} &\textbf{SHAP Value} &\textbf{Gene Name} &\textbf{SHAP Value} &\textbf{Gene Name} \\\midrule
Random &0.1167 &AT2G01100 &0.1436 &AT2G01100 &0.0289 &AT2G01100 \\
Forest &0.0804 &PDE331 &0.0981 &PDE331 &0.0183 &PAHX \\
&0.0325 &EXGT-A3 &0.041 &EXGT-A3 &0.0178 &EXGT-A3 \\\midrule
MLP &0.0098 &AT2G39855 &0.0106 &AT3G06880 &0.0119 &AT5G49920 \\
&0.0094 &BZIP28 &0.0103 &SNOR111 &0.0107 &AT5G09865 \\
&0.0093 &AT1G69630 &0.0099 &AT3G10210 &0.0103 &SQE6 \\\midrule
\textbf{HGNN} &> 1.0 &LOX2 &> 1.0 &GSTF10 &> 1.0 &AT4G21970 \\
\textbf{+ MLP} &> 1.0 &MEE11 &> 1.0 &SQN &> 1.0 &AT2G40710 \\
&> 1.0 &3BETAHSD/D2 &> 1.0 &PGY1 &> 1.0 &AT3G52660 \\
\bottomrule
\end{tabular}
\end{table}

Hypergraph regression results are given in Table \ref{tab:regression-metrics}. The most error-prone model, according to mean absolute error (MAE) and root mean square error (rMSE) is the combined MLP and hypergraph (HGNN) model. While introducing some biological context in the form of hyperedges, 
adding the hypergraph convolution layers to the MLP does not immediately lend itself well to the regression task. Incorporating the same structure into a VAE further improves results, but losses remain high, however, especially for test samples. Depending on the data split, the loss for the deep learning models may be high or low, possibly due to a failure to capture variance between the samples. While the most sound model appears to be the random forest regressor, of course, we do have a strong indication as to which genes or features are most important.

\subsection{Explainability Baselines} \label{subec:explainability-baselines}
The most explanatory genes from the cross-validation experiments are given in Tables \ref{tab:top-genes-12} and \ref{tab:top-genes-345}. For simplicity, we discuss the top three genes, whose descriptions we retrieved from a database using the gene information from our \textit{Arabidopsis} dataset to discover which genes, if any, have known functions. Table \ref{tab:top-genes-12} gives the most explanatory genes for the three regression-only baseline models with respect to solidity and inflorescence height. For random forest regression, it seems SHAP had difficulty discovering a diverse set of genes that could explain each of the five phenotypes. The top gene, named AT2G01100, like many genes, is not a gene commonly studied, and thus offers no further information other than encoding proteins. The third-place gene, called EXGT-A3, which affects all phenotypes but inflorescence, impacts leaf maturation and structure. The development of leaves certainly affects photosynthesis, and consequently, our phenotype parameters. The SHAP values for the MLP model with respect to inflorescence height returned two interesting genes, i.e., LHCA2 and PSAL. Both genes impact a protein called Photosystem I, which directly impacts photosynthesis. As expected, for the hypergraph (HGNN) and MLP model, the SHAP method had difficulties in computing results. As a result, all the ``top'' explanatory genes did not offer further information, i.e., these genes' impacts are not well understood. Table \ref{tab:top-genes-345} gives the explanatory genes for the spectrometry-based phenotype parameters. The top genes returned for all three of these parameters did not return any commonly-studied genes; most of these are protein-coding, while some are involved with RNA production, which may impact some processes downstream. 
In any case, we need efficient biologically-informed models and explanatory methods that consider these associations between genes and the higher-order pairings. 
It confirms current understanding of gene-gene and gene-phenotype connections and can lead to new connection discoveries.

\section{Conclusions} 

In this work, to foster and accelerate research for the G2P challenge, we introduce the \textit{Arabidopsis} Genomics-Phenomics (AGP) dataset--a curated multi-modal dataset for the model plant \textit{Arabidopsis thaliana} that links gene expression with heterogeneous phenotypic traits. While data collection in this field has improved over the years, datasets integrating transcriptomics and phenomics for the same organisms remain rare and computationally underutilized. We demonstrate that current models—including Random Forests, MLPs, and biologically-informed hypergraph neural networks—show varying strengths across regression and interpretability tasks. 
Our work establishes not only performance baselines but also interpretability baselines through SHAP analysis, revealing gene–trait relationships and limitations in existing explainability frameworks for hypergraphs. By releasing our dataset and methodology, we aim to catalyze further research in multi-modal graph learning, biological model interpretability, and scalable genome-wide prediction methods.
While downstream applications such as plant breeding will benefit greatly, we should, however, acknowledge the risks associated with knowing how genes influence traits, especially if that knowledge is transferred to, say, humans. 
Nevertheless, we believe AGP can serve both the plant biology and the broader machine learning communities working on graph-based and explainable AI.

\bibliography{bib/egbib}{}
\bibliographystyle{plain}





\newpage
\section*{NeurIPS Paper Checklist}

\begin{enumerate}

\item {\bf Claims}
    \item[] Question: Do the main claims made in the abstract and introduction accurately reflect the paper's contributions and scope?
    \item[] Answer: \answerYes{} 
    \item[] Justification: We have made a careful effort to align the claims made in the abstract and introduction with what is discussed in the rest of the manuscript.

\item {\bf Limitations}
    \item[] Question: Does the paper discuss the limitations of the work performed by the authors?
    \item[] Answer: \answerYes{} 
    \item[] Justification: Given that this work discusses difficult challenges from biology (G2P) and machine learning (explaninability), we discuss the limitations of our work--the data and the models used.

\item {\bf Theory assumptions and proofs}
    \item[] Question: For each theoretical result, does the paper provide the full set of assumptions and a complete (and correct) proof?
    \item[] Answer: \answerNo{} 
    \item[] Justification: Our work centers around introducing our dataset for connecting gene expression and phenotype measurements and exploring baseline results to lay the foundation for future works.

    \item {\bf Experimental result reproducibility}
    \item[] Question: Does the paper fully disclose all the information needed to reproduce the main experimental results of the paper to the extent that it affects the main claims and/or conclusions of the paper (regardless of whether the code and data are provided or not)?
    \item[] Answer: \answerYes{} 
    \item[] Justification: Yes. We provide the dataset and code for the reviewers to view and run (see next checklist item).

\item {\bf Open access to data and code}
    \item[] Question: Does the paper provide open access to the data and code, with sufficient instructions to faithfully reproduce the main experimental results, as described in supplemental material?
    \item[] Answer: \answerYes{} 
    \item[] Justification: The dataset is provided in a Kaggle dataset\footnote{https://kaggle.com/datasets/ae18c38ebba55da31e046c9a9d0b33033ae5c4e3c851bf2a64c544764e443478}, and the code to run our experiments from the main manuscript are provided in an anonymous GitHub repository \footnote{https://anonymous.4open.science/r/arabidopsis-genomics-phenomics-8EE5/}. 

\item {\bf Experimental setting/details}
    \item[] Question: Does the paper specify all the training and test details (e.g., data splits, hyperparameters, how they were chosen, type of optimizer, etc.) necessary to understand the results?
    \item[] Answer: \answerYes{} 
    \item[] Justification: We provide experimental details and results in Section \ref{sec:experiments}.

\item {\bf Experiment statistical significance}
    \item[] Question: Does the paper report error bars suitably and correctly defined or other appropriate information about the statistical significance of the experiments?
    \item[] Answer: \answerYes{} 
    \item[] Justification: While our experiments do not test statistical significance, we report the deviation of the regression baselines in Section \ref{subsec:regression-experiments}.

\item {\bf Experiments compute resources}
    \item[] Question: For each experiment, does the paper provide sufficient information on the computer resources (type of compute workers, memory, time of execution) needed to reproduce the experiments?
    \item[] Answer: \answerYes{} 
    \item[] Justification: We provide experimental setup information in Section \ref{subsec:experiment-setup}, and any further details are given in the supplementary material.
    
\item {\bf Code of ethics}
    \item[] Question: Does the research conducted in the paper conform, in every respect, with the NeurIPS Code of Ethics \url{https://neurips.cc/public/EthicsGuidelines}?
    \item[] Answer: \answerYes{} 
    \item[] Justification: We believe that this dataset alone does not violate ethics principles set forth by NeurIPS. In the future, however, it may contribute to ethics issues such as the dual use problem and discrimination of individuals on the basis of their genes influencing personal traits.

\item {\bf Broader impacts}
    \item[] Question: Does the paper discuss both potential positive societal impacts and negative societal impacts of the work performed?
    \item[] Answer: \answerYes{} 
    \item[] Justification: From Sections \ref{sec:intro} and \ref{sec:background-related-work} we discuss the dataset and computational limitations AGP addresses. Fostering further research into accurately mapping genes to traits will impact many fields in agriculture and medicine. At the same time, we also acknowledge the potential negative impacts knowing exactly how genes and traits are linked for the future.
    
\item {\bf Safeguards}
    \item[] Question: Does the paper describe safeguards that have been put in place for responsible release of data or models that have a high risk for misuse (e.g., pretrained language models, image generators, or scraped datasets)?
    \item[] Answer: \answerNA{} 
    \item[] Justification: The core of the AGP dataset is comprised of pictures and measurements with respect to plant specimens. There is a rather low risk of misuse for harm to other people. The AGP dataset alone, like other publicly-available datasets, cannot pose such risks that language models, image generators, etc., do.

\item {\bf Licenses for existing assets}
    \item[] Question: Are the creators or original owners of assets (e.g., code, data, models), used in the paper, properly credited and are the license and terms of use explicitly mentioned and properly respected?
    \item[] Answer: \answerYes{} 
    \item[] Justification: The gene expression and phenotype measurements are our original assets. We properly provide credit for data that is not ours in Section \ref{sec:dataset} and the supplementary materials.

\item {\bf New assets}
    \item[] Question: Are new assets introduced in the paper well documented and is the documentation provided alongside the assets?
    \item[] Answer: \answerYes{} 
    \item[] Justification: We provide an overview of the new AGP dataset in the manuscript. The supplementary materials provide further details. The data itself also will contain documentation (i.e., a description of what information each file has).

\item {\bf Crowdsourcing and research with human subjects}
    \item[] Question: For crowdsourcing experiments and research with human subjects, does the paper include the full text of instructions given to participants and screenshots, if applicable, as well as details about compensation (if any)? 
    \item[] Answer: \answerNA{} 
    \item[] Justification: The AGP dataset and methods used in this work do not involve any human subject information. The gene and GO term (biological function) information was gathered and prepared by experts in the biological sciences, and thus is not crowdsourced from anonymous sources. This information was made publicly available in different sites across the Internet specifically for research purposes.

\item {\bf Institutional review board (IRB) approvals or equivalent for research with human subjects}
    \item[] Question: Does the paper describe potential risks incurred by study participants, whether such risks were disclosed to the subjects, and whether Institutional Review Board (IRB) approvals (or an equivalent approval/review based on the requirements of your country or institution) were obtained?
    \item[] Answer: \answerNA{} 
    \item[] Justification: There are no study participants or any personal information contained in our dataset. Any information only pertains to plant specimens of the species \textit{Arabidopsis thaliana}.

\item {\bf Declaration of LLM usage}
    \item[] Question: Does the paper describe the usage of LLMs if it is an important, original, or non-standard component of the core methods in this research? Note that if the LLM is used only for writing, editing, or formatting purposes and does not impact the core methodology, scientific rigorousness, or originality of the research, declaration is not required.
    \item[] Answer: \answerNo{} 
    \item[] Justification: We do not use LLMs in any capacity of the core method of this work or the AGP dataset. We do note, however, the connection between graphs and language models in some recent works, making a connection possible in the future, but that is beyond the scope of our work (Section \ref{sec:background-related-work} contains some citations).

\end{enumerate}

\end{document}